\documentclass[11pt,twoside,a4paper]{article}

\usepackage{graphicx}
\usepackage{dcolumn}
\usepackage{bm}
\usepackage{units} 
\usepackage{color}

\usepackage[utf8]{inputenc}
\usepackage[T1]{fontenc}
\usepackage{mathptmx}
\usepackage{amsmath}
\usepackage[numbers,sort&compress]{natbib}
\usepackage{authblk}

\begin{document}

\title{Mode-locking in vertical external-cavity surface-emitting lasers with type-II quantum-well configurations}

\author[1]{I.~Kilen}
\author[1,2]{S.~W.~Koch}
\author[1,3]{J.~Hader}
\author[1,3,4]{J.~V.~Moloney}
\affil[1]{College of Optical Sciences, University of Arizona, 1630 East University Boulevard, Tucson, AZ 85721, USA}
\affil[2]{Department of Physics and Material Sciences Center, Philipps-Universit\"at Marburg, Renthof~5, 35032 Marburg, Germany}
\affil[3]{Nonlinear Control Strategies Inc., Tucson, AZ, 85704, USA}
\affil[4]{Department of Mathematics, University of Arizona, 617 N. Santa Rita Ave., Tucson, AZ 85721, USA}
\setcounter{Maxaffil}{0}
\renewcommand\Affilfont{\itshape\small}

\maketitle

\begin{abstract}
A microscopic study of mode-locked pulse generation is presented for vertical external-cavity surface-emitting lasers utilizing type-II quantum well configurations. The coupled Maxwell semiconductor Bloch equations are solved numerically where the type-II carrier replenishment is modeled via suitably chosen reservoirs. Conditions for stable mode-locked pulses are identified allowing for pulses in the \unit[100]{fs} range. Design strategies for type-II configurations are proposed that avoid potentially unstable pulse dynamics.
\end{abstract}

\section{Introduction}

Vertical external-cavity surface-emitting lasers (VECSELs) are highly versatile, relatively low-cost semiconductor disk lasers capable of continuous wave (CW) and mode-locked operation\cite{KellerTropper06, tropper2012ultrafast, tilma2015recent, rahimi2016recent}. Up to now, most systems are based on quantum well (QW) heterostructures with type-I band alignment\cite{kroemer1996band}, i.e. configurations where the optical transitions involve electrons and holes in the same QW. Here, the highest peak powers for multi-mode CW output has been reported as \unit[106]{W} \cite{ElectronLett12, LaserPhotonRev12}, single mode-operation has yielded \unit[23]{W} with a distributed Bragg reflector (DBR) \cite{zhang201423}, and \unit[16]{W} without a DBR\cite{yang201816}. In mode-locked operation, pulses with temporal durations of about \unit[100]{fs}\cite{waldburger2016high, klopp13} and below\cite{laurain2018modeling} have been realized.

In modern heterostructure growth, it is also possible to realize systems with a type-II band alignment, where the energetically lowest conduction band minima and the highest valence band maxima are in different QWs \cite{kroemer1996band}. 
Despite the fact that the optical transitions are spatially indirect in this case, rather large optical gain can be realized rivaling that of traditional type-I systems\cite{buckers2008microscopic}. Due to the added flexibility to separately optimize the two QW configurations involved in the type-II transition, one can realized optical transitions in a wider frequency range than possible with type-I systems. Moreover, so-called type-II W-configurations have been designed for long-wavelength operation with the goal to reduce the intrinsic Auger losses and simultaneously maintaining high spectral amplification\cite{meyer1995type, zegrya1996theory}. 

Recently, W-well based VECSELs for operation in the \unit[1200]{nm} wavelength regime have been realized and shown to reach \unit[4]{W} of multi-mode and \unit[0.35]{mW} of single-mode CW output power\cite{moller2016type, moller2016fundamental}. In pulsed operation, electrical injection pumped edge-emitting gain chips based on type-II QW design have produced pulses with \unit[1.4]{W} output power per facet\cite{fuchs2016electrical}. Not only has it been demonstrated that type-II QWs are able to reach new wavelength ranges but the studies of the systems in the \unit[1.3]{$\mu$m} region also showed that the type-II QW configuration tends to be less sensitive to changes in temperature\cite{fuchs2018high}.

On the basis of these promising results, in particular in view of the relatively wide flexibility to choose the optical transition frequency, it is interesting to investigate to which degree type-II W-laser configurations can be used for mode-locking and ultra short pulse generation. For optical pumping into energetically high lying barrier states, it is very likely that the resulting, spatially separated densities of electrons and holes in the active lasing transition are somewhat different due to their intrinsically different relaxation processes. In mode-locked operation, the propagating intracavity pulse is amplified in the gain chip QWs and partially absorbed by the semiconductor saturable absorber mirror (SESAM). In order to obtain stable mode-locked pulses, it is mandatory that both, the gain chip and the SESAM recover to the same state during every round-trip. 

In comparison to type-I systems, the type-II QWs exhibit a somewhat different carrier recovery dynamics such that it is not immediately clear if one can match the gain chip and absorber dynamics allowing for mode-locking of the devices. Recent experimental measurements show that type-II QW gain recovery is slower than that of type-I QWs\cite{lammers2016gain}, which could adversely influence the pulse energy and mode-locking dynamics. Since the amount of amplification/absorption is nonlinearly dependent on the instantaneous QW carrier state, one has to systematically account for the microscopic QW carrier dynamics and couple this back to the pulse propagation in order to quantitatively model the mode-locking dynamics in type-II QWs.

In this paper, we investigate the potential for stable mode-locked VECSELs based on type-II QWs. For this purpose, we numerically solve Maxwell's equation for pulse propagation through the VECSEL coupled to the semiconductor Bloch equations (SBE) that describe the macroscopic polarization of each QW. To keep the approach as simple as possible, we model the microscopic QW polarization dephasing and the carrier scattering with effective rates where the polarization relaxes to zero and the carriers scatter towards an instantaneous quasi-Fermi distribution\cite{kilen2016fully}. In the past, we used this scheme to analyze mode-locking in VECSELs with type-I QWs and successfully compared the outcome to experimental results such that, with the appropriate modification introduced here, this microscopically based model should be well suited for testing the stability of mode-locking in type-II QWs. 

\section{Theory}

\begin{figure}[ht]
\centerline{\includegraphics[width=0.5\linewidth]{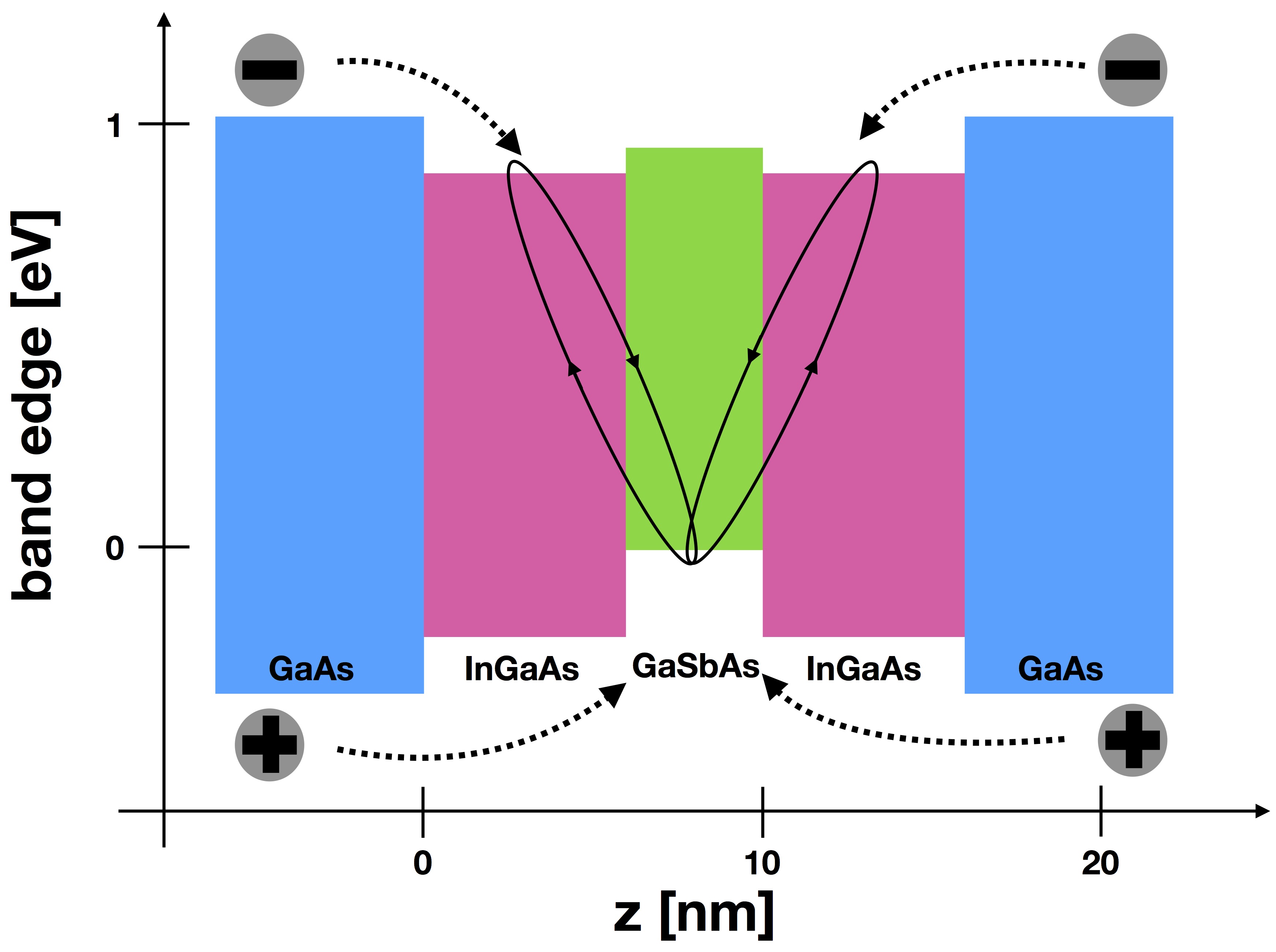}}
\caption{A schematic of the spatial design of the type-II QW potentials with GaAs (blue), InGaAs (magenta), and GaSbAs (green) material layers. The solid arrows indicate the lasing transition simulated using the SBE.}
\label{fig:structureA}
\end{figure}

We consider the generic structure configuration shown in Fig.~\ref{fig:structureA}. Here, the optical pump is absorbed in the GaAs barrier material and generates carriers which subsequently relax from the spatially delocalized initial states into the energetically lower confined QW states via carrier-carrier and carrier-phonon scattering. This carrier capture process has been modeled microscopically with good agreement to experimental observations\cite{hader2004structural}. During and after the relaxation into the lower energy states, the carriers eventually equilibrate energetically close to the lasing transition. 

Since the detailed, fully microscopic many-body calculation of the high energy carriers relaxing into the lower lasing states is computationally expensive, we introduce a simplified treatment that still allows us to capture the main features of a type-II W-laser configuration. The constant background of carriers, available to the QWs from barrier pump absorption, is modeled as a static Fermi distribution. These background carriers scatter into a QW high-energy reservoir state which will dynamically release carriers into the lower-energy lasing transition. Due to the different well configurations for electrons and holes in a type-II QW, the rate of transfer of carriers from the reservoir to the lasing transition may differ for electrons and holes, which in general will result in unequal densities in the active transition.

In our approach, the relaxation of carriers from the constant, pump induced background distributions, $\mathrm{F}^{(\text{e}/\text{h})}_\text{k}$, to the QW reservoir is modeled using a scattering time $\tau_\text{scatt}$. The carriers from the reservoir are assumed to scatter into the lasing transition with a characteristic electron (hole) time $\tau_{e}$ ($\tau_{h}$). This relaxation is modeled using $-(\mathrm{n}^{(\text{e}/\text{h})}_\text{k} -\mathrm{n}^{(\text{e}/\text{h}),\text{res}}_\text{k})/\tau_{(\text{e}/\text{h})}$. 
\begin{figure}[ht]
\centerline{\includegraphics[width=0.5\linewidth]{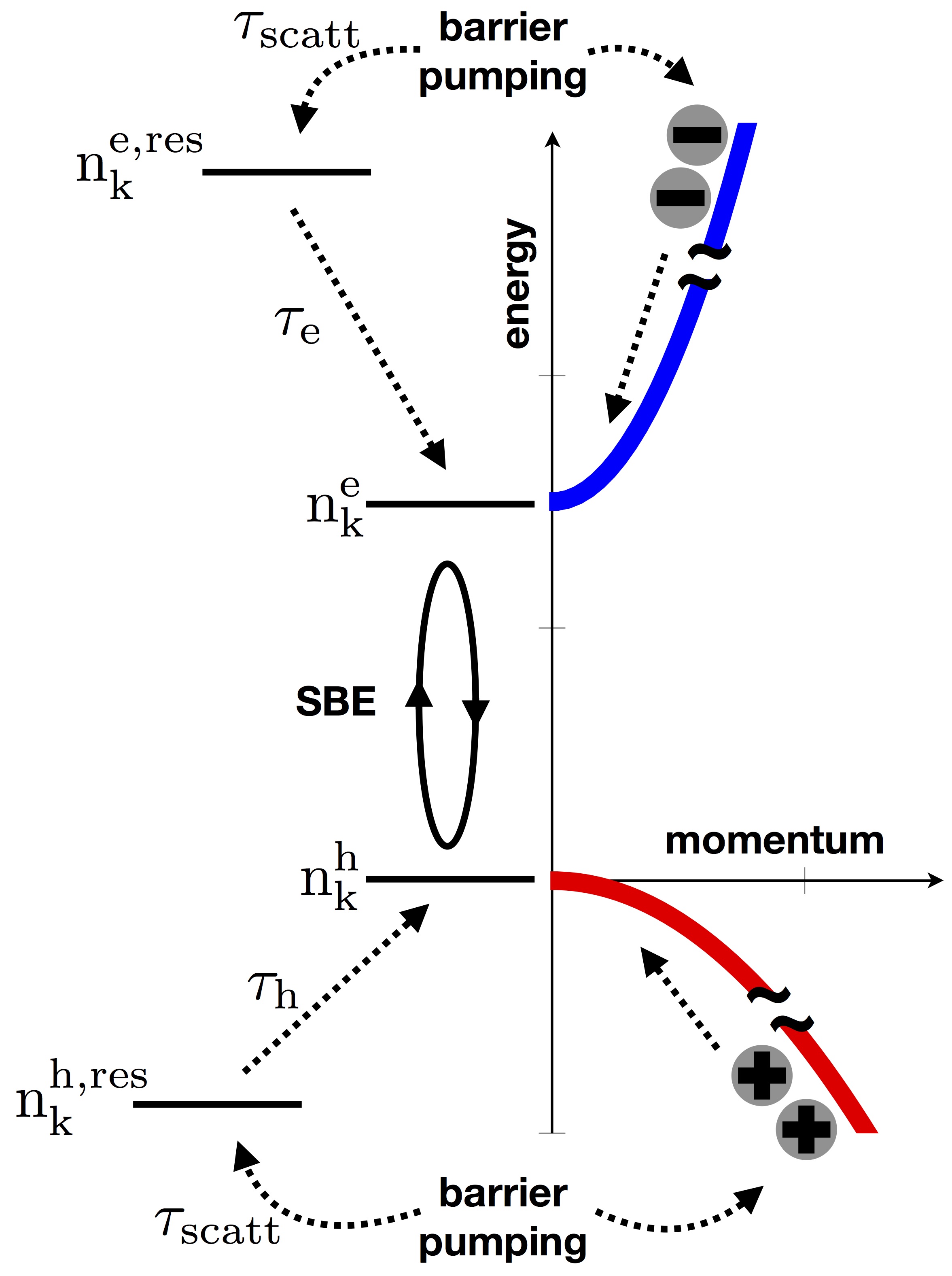}}
\caption{A diagram showing the relaxation of barrier pumped carriers in a type-II QW and the important model parameters: the momentum resolved high energy reservoir carriers, $\mathrm{n}_\text{k}^{(\text{e}/\text{h}),\text{res}}$, that scatter to lower energy states and replenish the active lasing carriers, $\mathrm{n}_\text{k}^{(\text{e}/\text{h})}$, on a timescale $\tau_{(\text{e}/\text{h})}$. Barrier pumped carriers relax into the reservoir at a timescale, $\tau_\text{scatt}$. The dashed arrows indicate transitions that involve carrier scattering and the solid arrows is the lasing transition simulated using the SBE.}
\label{fig:structureB}
\end{figure}
Fig.~\ref{fig:structureB} contains a schematic of this relaxation model. Following this, the reservoir electrons relax according to, $\text{d}/\text{dt}\, \mathrm{n}^{\text{e},\text{res}}_\text{k}(t) = -(\mathrm{n}^{\text{e},\text{res}}_\text{k} - \mathrm{F}^\text{e}_\text{k})/\tau_\text{scatt} + (\mathrm{n}^{\text{e}}_\text{k} -\mathrm{n}^{\text{e},\text{res}}_\text{k})/\tau_{\text{e}}$, with a corresponding equation for the holes. 

With regards to the numerical values for the various relaxation times, we note that in experimentally realized heterostructures, the capture times depend on the details of the entire system which define the multitude of energy bands that characterize carriers in continuum (barrier states) and captured configurations in the QW. For barrier pumping, the microscopic mechanism of the capture process is a combination of Coulomb carrier-carrier and carrier-phonon scattering. The resulting capture rates will thus depend on the precise multi-band configuration as well as on the energy and intensity of the pump. Using our fully microscopic analysis for a characteristic configuration (see supplementary material), we find timescales on the order of ten picoseconds. Since the goal of this paper is a feasibility study for mode-locking in type-II W-laser configurations, we thus perform studies for a range of realistic capture rates in the range of \unit[5]{ps} to \unit[40]{ps} and identify their influence on the emitted pulse properties.

Before we proceed to a presentation of our numerical results, we want to note that one feature of our pump model is that the replenishment of the carrier density in the active lasing transition is limited by the reservoir carrier density level. During lasing operation, carriers are scattered into the reservoir on the timescale of $\tau_\text{scatt}$ and lost to the lasing transition on the timescale of $\tau_{(\text{e}/\text{h})}$. On this basis, the resulting dynamic interplay between carriers in the active lasing transition and reservoir produces two effective timescales for the carrier density recovery of the active lasing transition, $\text{T}^{(\text{e}/\text{h})}_\pm$, that are nonlinear combinations of the scattering times, $\tau_\text{e}$, $\tau_\text{h}$, and $\tau_\text{scatt}$. Details on the derivation and analytic expression for these effective timescales in the absence of an electric field are found in the supplementary material. In particular, we identify a timescale for the slow recovery of carriers, $\text{T}^{(\text{e}/\text{h})}_- \geq 2\tau_\text{scatt}$, and this inequality becomes larger for increasing $\tau_{(\text{e}/\text{h})}$. The fast carrier recovery time satisfies, $0<\text{T}^{(\text{e}/\text{h})}_+ < \tau_\text{scatt}$, and approaches zero for shorter scattering times, $\tau_{(\text{e}/\text{h})}$. Note that, as a consequence of the two-stage relaxation model presented above, type-II QWs will have a slower gain recovery rate than type-I QWs using only, $\tau_\text{scatt}$, as a characteristic timescale for carrier recovery. 

\section{Results}

To illustrate the generic similarities and differences, we analyze the influence of the type-II QW carrier relaxation on mode-locked pulses via a comparison to the results for the same VECSEL setup with type-I QWs\cite{kilen2018vecsel}. We choose a configuration where the gain chip consists of a DBR with 30 pairs of AlGaAs, an active region containing the QWs in a resonant periodic gain arrangement, an InGaP cap layer, and a dispersion compensating coating from $\mathrm{Ta}_2\mathrm{O}_5$/$\mathrm{SiO}_2$. In this design, 10 QWs are arranged on the antinodes of a standing wave from the DBR centered at \unit[1030]{nm}. The external cavity is configured with a simple SESAM on top of a \unit[1]{\%} output coupling mirror separated from the gain chip by an air gap that results in a pulse round-trip time of \unit[21]{ps}. For the numerical investigations presented below, we use $\tau_\text{scatt}=$  \unit[30]{ps}, a lattice temperature of \unit[300]{K} and a constant background density of \unit[$1.82\cdot 10^{12}$]{$\mathrm{cm}^{-2}$}. As a comparison, the same VECSEL cavity with type-I QWs relaxing to the same background Fermi distribution with a single characteristic timescale, $\tau_\text{scatt}$, produced a mode-locked pulse with a full-width half-maximum (FWHM) of \unit[102]{fs} and an output peak intensity of \unit[0.16]{$\text{MW}/\text{cm}^2$}.

\begin{figure}[ht]
\centerline{\includegraphics[width=0.6\linewidth]{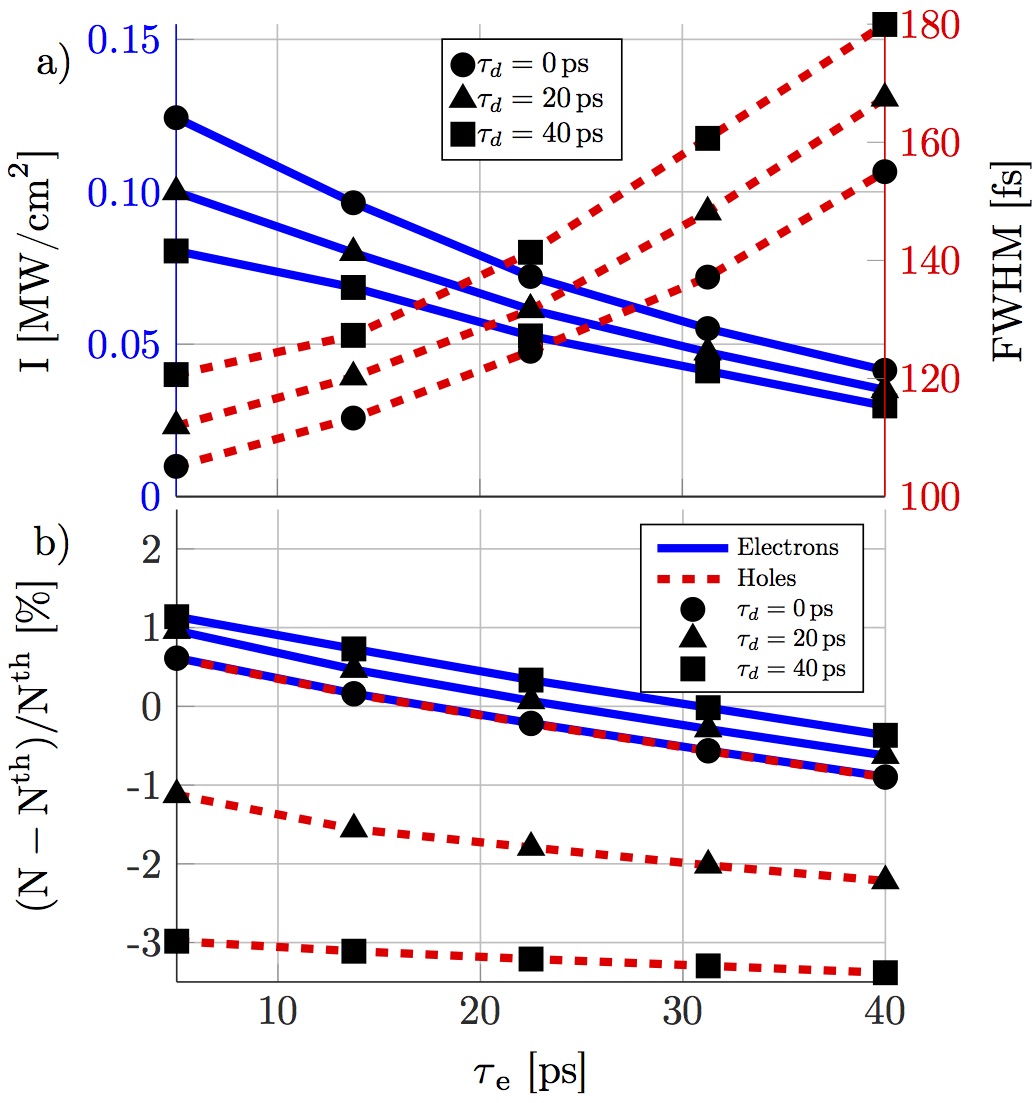}}
\caption{A summary of mode-locked pulses from type-II QWs for electron scattering times, $\tau_\text{e}$, with select hole rates given by, $\tau_\text{h} = \tau_\text{e}+\tau_\text{d}$, for $\tau_\text{d} = $ \unit[0]{ps} (circles), \unit[20]{ps} (triangles), and \unit[40]{ps} (squares).  a) The resulting mode-locked output peak intensity (blue) and FWHM (red). b) The maximal QW carrier density during a round-trip of mode-locked operation relative to the threshold density $\text{N}^\text{th}$.}
\label{fig:overview}
\end{figure}

Figure \ref{fig:overview}a) gives an overview of stable mode-locked pulse peak intensities and temporal FWHM, produced with type-II QWs, for multiple different electron and hole scattering times. Here, we have sampled the hole scattering times according to, $\tau_\text{h} = \tau_\text{e}+\tau_\text{d}$, where $\tau_\text{d} = $ \unit[0]{ps} (circles), \unit[20]{ps} (triangles), and \unit[40]{ps} (squares). For all values for $\tau_\text{d}$ examined here,  the pulse peak intensity increases by a factor of about 3 and the FWHM decreases by about \unit[66]{\%} when the electron scattering time, $\tau_\text{e}$, is reduced from \unit[40]{ps} to \unit[5]{ps}. On the other hand, increasing the hole scattering time by, $\tau_\text{d}$, with a fixed, $\tau_\text{e}$, results in a \unit[28-35]{\%} decrease in pulse peak intensity and a \unit[15-16]{\%} increase in pulse FWHM. As expected, a lower scattering time results in faster carrier replenishing and more carriers in the lasing transition and, thus, support a higher pulse peak intensity. In contrast, a longer scattering time will reduce the QW gain and result in a higher pulse temporal duration. Overall, the pulse fluence decreases for longer carrier relaxation times because the intensity drops faster than the pulse widens. For the range of $\tau_\text{e}$ examined here, the fluence drops by about a factor of 2 for a fixed $\tau_\text{d}$ and, for a fixed $\tau_\text{e}$, decreases by about \unit[17-20]{\%} with varying $\tau_\text{d}$. 

In Fig.~\ref{fig:overview}b), we have recorded the QW carrier density immediately before a mode-locked pulse extracts carriers. The densities are relative to the threshold density, $\text{N}^\text{th}\approx$\unit[1.739]{$\cdot 10^{16}\text{m}^{-2}$}, where the spectral amplification of the gain chip is equal to the cavity loss when the gain chip QW carriers are assumed to be in equilibrium at the background pump density. There is a \unit[0.4-1.6]{\%} increase in carrier density for fixed, $\tau_\text{d}$, when reducing the electron scattering time over this plot range. For fixed, $\tau_\text{e}$, the increased time difference, $\tau_\text{d}$,  results in about a \unit[0.5]{\%} increase in electron density and a \unit[2.5-3.6]{\%} decrease in hole density. We expected the reduced hole density because, $\tau_\text{h}$, is increasing, which results in a slower transfer of carriers from the reservoir into the lasing transition. It is somewhat unexpected that the electron density slightly increases. This happens in order to compensate for the reduced hole density. The latter change is minimal and the average density decreases by about \unit[1-2.5]{\%}.

Figure \ref{fig:overview}b) shows that for longer electron scattering times the non-equilibrium carrier density goes slightly below the threshold density, $\text{N}^\text{th}$, for both, electrons and holes. The threshold density is common in characterization of lasers and is significant because spontaneous emission cannot grow below this density. However, mode-locked pulses can operate stably below the threshold density\cite{kilen2017non} because non-equilibrium pulse amplification is obviously different from CW operation and, $\text{N}^\text{th}$, is computed while QW carriers are in equilibrium Fermi distributions. In our simulations, the pulse is first established using the background pump density \unit[$1.82\cdot 10^{16}$]{$\mathrm{m}^{-2}$}, which provides net amplification inside the VECSEL.

In Fig.~\ref{fig:overview}, we see that the mode-locked pulse peak intensity depends critically on $\tau_{(\text{e}/\text{h})}$. The mode-locked pulses from type-II QWs are both longer and have lower peak intensities than the corresponding type-I QWs. A longer electron and hole relaxation time will in general result in lower mode-locked pulse energy. We have not attempted to optimize the type-II QWs for ultrashort pulse generation or high peak intensity at this time. However, the results show that a shorter electron/hole relaxation time will result in more carriers in the lasing transition, which will give shorter FWHM and higher peak intensity.

When comparing Figs.~\ref{fig:overview}a) and b) over the given range of electron scattering times, we observe that the pulse fluence decreases by a factor of 2, while the total carrier density only decreases by \unit[1-2.5]{\%}. It is not surprising that very similar carrier densities can produce very different mode-locked pulses because the total carrier density is not directly available to the pulse itself. Instead, the pulse can only extract energy from a subset of carriers in each QW, namely the inverted carriers\cite{kilen2014ultrafast}. For type-II QWs, the principal difference is that the electron and hole occupation numbers will relax differently based on their respective relaxation times and results in unequal electron and hole densities. However, the pulse still interacts with the QW inversion, $n^\text{e}_k + n^\text{h}_k -1$, that depends on the momentum resolved carrier occupation numbers.

\begin{figure}[ht]
\centerline{\includegraphics[width=0.6\linewidth]{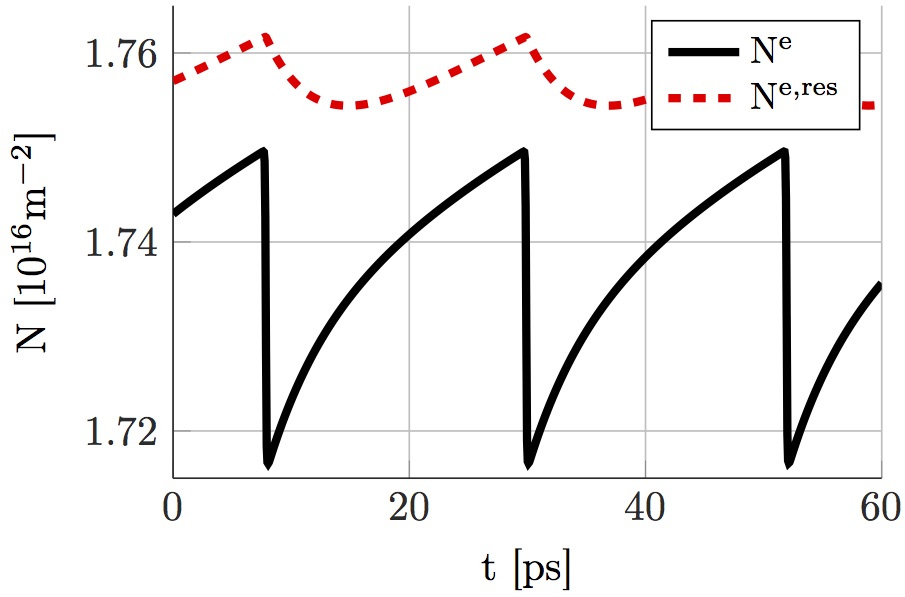}}
\caption{Total density of electrons in the reservoir and in the active transition  during stable mode-locked operation. In this example: ${\tau_e = 10\,\text{ps}}$ and ${\tau_\text{h} = 25\,\text{ps}}$.}
\label{fig:modeLockedExample}
\end{figure}
Our detailed analysis indicates that the reservoir relaxation and the associated electron and hole relaxation times can become a source for instabilities in the mode-locking dynamics. To to see this, we first consider a typical example with a stable mode-locked pulse. Fig.~\ref{fig:modeLockedExample} shows an example of the total density of reservoir electrons and active lasing electrons during stable mode-locked operation. Note, that the drop of carriers is due to the propagating mode-locked pulse extracting carriers each round-trip. A stable mode-locked pulse appears when the reservoir maintains a sufficient reserve of carriers and the lasing transition has enough carriers to support the propagating mode-locked pulse. This case was examined in detail in Fig.~\ref{fig:overview}. Another limit to consider is when the electron/hole relaxation times become very long compared to the reservoir relaxation time. In this limit, there are not enough carriers in the lasing transition to produce a mode-locked pulse with sufficient energy to bleach the SESAM, which will prevent the formation of a stable mode-locked pulse. 

A true non-equilibrium pulse instability can appear when one or both of the electron/hole relaxation times are sufficiently short. In this case, too many carriers relax into the lasing transition during interaction with the single mode-locked pulse, and the unsaturated carriers attempt to produce a second pulse immediately following the first. This case is similar to a previously studied situation where multiple cavity pulses were created because of too high QW carrier density \cite{kilen2014ultrafast}. The main difference is that, the current unstable pulse form does not break apart into multiple pulses. The tail behind the leading pulse will, over thousands of round-trips, grow into a secondary pulse that extracts energy from a spectral region near the carrier reservoir of the original pulse. The system attempts to accommodate both carrier reservoirs, but is unable to find a balance that supports two pulses. The leading pulse is suppressed in favor of the secondary pulse and this process repeats by creating a tail behind the new pulse. This situation can be avoided and stable mode-locking can be established by reducing the density of the background barrier distribution, $\mathrm{F}^{(\text{e}/\text{h})}_\text{k}$, i.e., by reducing the pump. The supplementary material contains more details on this instability.

Altogether, we identify three stability regimes for mode-locked pulses that can be characterized by looking at the reservoir carriers. The mode-locked pulse is stable when the reservoir has a sufficient carrier density to replenish the active lasing transition, i.e., the electron/hole relaxation times are not too short and thus allow for a balance between the reservoir and lasing transition carriers. Unstable lasing occurs when the reservoir carriers relax too quickly into the lasing transition during interaction with the pulse. This fast recovery can produce unsaturated carriers that in turn can cause an additional pulse to appear in a time window after the original cavity pulse. Finally, too long electron/hole relaxation times result in no mode-locked pulse in the cavity.

\section{Conclusion}

In summary, we presented numerical results analyzing the feasibility to generate short pulse mode-locking in type-II QW W-laser configurations. We predict stable mode-locked pulses with temporal durations slightly above \unit[100]{fs} and peak intensities \unit[0.1]{MW/cm$^2$} range. We find stable mode-locked pulses with somewhat lower peak intensities and longer FWHM than the corresponding VECSEL with type-I QWs. Whereas the pulse details depend on the exact system parameters, the overall mode-locked pulse behavior is rather robust against small changes in the exact carrier capture dynamics. We derived analytical expressions that should help with the design of type-II QW based gain chips, where the type-II QW configuration allows for increased flexibility to chose the central pulse emission wavelength.

See \emph{supplementary material} for the derivation of the carrier recovery timescales, $\text{T}_\pm$, details on the unstable mode-locked pulse, and an example many-body calculation of carriers scattering into the type-II QW.
 
This material is based upon work supported by the Air Force Office of Scientific Research under award numbers FA9550-17-1-0246. The Marburg research is supported by the Deutsche Forschungsgemeinschaft via the Sonderforschungsbereich 1083. We thank Wolfgang Stolz for insightful discussions.

\bibliographystyle{plainnat}
\bibliography{main}

\begin{thebibliography}{25}
\providecommand{\natexlab}[1]{#1}
\providecommand{\url}[1]{\texttt{#1}}
\expandafter\ifx\csname urlstyle\endcsname\relax
  \providecommand{\doi}[1]{doi: #1}\else
  \providecommand{\doi}{doi: \begingroup \urlstyle{rm}\Url}\fi

\bibitem[B{\"u}ckers et~al.(2008)B{\"u}ckers, Thr{\"a}nhardt, Koch, Rattunde,
  Schulz, Wagner, Hader, and Moloney]{buckers2008microscopic}
C~B{\"u}ckers, A~Thr{\"a}nhardt, Stephan~W Koch, M~Rattunde, N~Schulz,
  J~Wagner, Jorg Hader, and Jerome~V Moloney.
\newblock Microscopic calculation and measurement of the laser gain in a
  {(GaIn)Sb} quantum well structure.
\newblock \emph{Appl. Phys. Lett.}, 92\penalty0 (7):\penalty0 071107, 2008.

\bibitem[Fuchs et~al.(2016)Fuchs, Berger, M{\"o}ller, Weseloh, Reinhard, Hader,
  Moloney, Koch, and Stolz]{fuchs2016electrical}
C~Fuchs, C~Berger, C~M{\"o}ller, M~Weseloh, S~Reinhard, Jorg Hader, Jerome~V
  Moloney, Stephan~W Koch, and W~Stolz.
\newblock Electrical injection type-{II} {(GaIn) As/Ga (AsSb)/(GaIn) As} single
  {'W'}-quantum well laser at \unit[1.2]{$\mu$m}.
\newblock \emph{Electron. Lett.}, 52\penalty0 (22):\penalty0 1875--1877, 2016.

\bibitem[Fuchs et~al.(2018)Fuchs, Br{\"u}ggemann, Weseloh, Berger, M{\"o}ller,
  Reinhard, Hader, Moloney, B{\"a}umner, Koch, et~al.]{fuchs2018high}
C~Fuchs, A~Br{\"u}ggemann, MJ~Weseloh, C~Berger, C~M{\"o}ller, S~Reinhard,
  J~Hader, JV~Moloney, A~B{\"a}umner, SW~Koch, et~al.
\newblock High-temperature operation of electrical injection type-{II}
  {(GaIn)As/Ga (AsSb)/(GaIn)} as "{W}"-quantum well lasers emitting at
  \unit[1.3]{$\mu$m}.
\newblock \emph{Sci. Rep.}, 8\penalty0 (1):\penalty0 1422, 2018.

\bibitem[Hader et~al.(2004)Hader, Moloney, and Koch]{hader2004structural}
Jorg Hader, Jerome~V Moloney, and Stephan~W Koch.
\newblock Structural dependence of carrier capture time in semiconductor
  quantum-well lasers.
\newblock \emph{Appl. Phys. Lett.}, 85\penalty0 (3):\penalty0 369--371, 2004.

\bibitem[Heinen et~al.(2012)Heinen, Wang, Sparenberg, Weber, Kunert, Hader,
  Koch, Moloney, Koch, and Stolz]{ElectronLett12}
B.~Heinen, T.-L. Wang, M.~Sparenberg, A.~Weber, B.~Kunert, J.~Hader, S.~W.
  Koch, J.~V. Moloney, M.~Koch, and W.~Stolz.
\newblock \unit[106]{W} continuous-wave output power from
  vertical-external-cavity surface-emitting laser.
\newblock \emph{Electron. Lett.}, 48\penalty0 (9):\penalty0 516, 2012.

\bibitem[Keller and Tropper(2006)]{KellerTropper06}
U.~Keller and A.~C. Tropper.
\newblock Passively modelocked surface-emitting semiconductor lasers.
\newblock \emph{Phys. Rep.}, 429:\penalty0 67--120, 2006.

\bibitem[Kilen et~al.(2014)Kilen, Hader, Moloney, and Koch]{kilen2014ultrafast}
I.~Kilen, Jorg Hader, Jerome~V. Moloney, and Stephan~W. Koch.
\newblock Ultrafast nonequilibrium carrier dynamics in semiconductor laser mode
  locking.
\newblock \emph{Optica}, 1\penalty0 (4):\penalty0 192--197, 2014.

\bibitem[Kilen et~al.(2016)Kilen, Koch, Hader, and Moloney]{kilen2016fully}
I.~Kilen, S.~W. Koch, J.~Hader, and J.~V. Moloney.
\newblock Fully microscopic modeling of mode locking in microcavity lasers.
\newblock \emph{JOSA B}, 33\penalty0 (1):\penalty0 75--80, 2016.

\bibitem[Kilen et~al.(2017)Kilen, Koch, Hader, and Moloney]{kilen2017non}
I.~Kilen, S.~W. Koch, J.~Hader, and J.~V. Moloney.
\newblock Non-equilibrium ultrashort pulse generation strategies in {VECSELs}.
\newblock \emph{Optica}, 4\penalty0 (4):\penalty0 412--417, 2017.

\bibitem[Kilen et~al.(2018)Kilen, Koch, Hader, and Moloney]{kilen2018vecsel}
I.~Kilen, S.~W. Koch, J.~Hader, and J.~V. Moloney.
\newblock {VECSEL} design for high peak power ultrashort mode-locked operation.
\newblock \emph{Appl. Phys. Lett.}, 112\penalty0 (26):\penalty0 262105, 2018.

\bibitem[Klopp et~al.(2011)Klopp, Griebner, Zorn, and Weyers]{klopp13}
P.~Klopp, U.~Griebner, M.~Zorn, and M.~Weyers.
\newblock Pulse repetition rate up to \unit[92]{GHz} or pulse duration shorter
  than \unit[110]{fs} from a mode-locked semiconductor disk laser.
\newblock \emph{Appl. Phys. Lett.}, 98:\penalty0 071103, 2011.

\bibitem[Kroemer(1996)]{kroemer1996band}
Herbert Kroemer.
\newblock Band offsets and chemical bonding: the basis for heterostructure
  applications.
\newblock \emph{Physica Scripta}, 1996\penalty0 (T68):\penalty0 10, 1996.

\bibitem[Lammers et~al.(2016)Lammers, Stein, Berger, M{\"o}ller, Fuchs,
  Ruiz~Perez, Rahimi-Iman, Hader, Moloney, Stolz, et~al.]{lammers2016gain}
Christian Lammers, Markus Stein, Christian Berger, Christoph M{\"o}ller,
  Christian Fuchs, A~Ruiz~Perez, Arash Rahimi-Iman, J{\"o}rg Hader, JV~Moloney,
  Wolfgang Stolz, et~al.
\newblock Gain spectroscopy of a type-{II} {VECSEL} chip.
\newblock \emph{Appl. Phys. Lett.}, 109\penalty0 (23):\penalty0 232107, 2016.

\bibitem[Laurain et~al.(2018)Laurain, Kilen, Hader, Ruiz~Perez, Ludewig, Stolz,
  Addamane, Balakrishnan, Koch, and Moloney]{laurain2018modeling}
Alexandre Laurain, Isak Kilen, Jorg Hader, Antje Ruiz~Perez, Peter Ludewig,
  Wolfgang Stolz, Sadhvikas Addamane, Ganesh Balakrishnan, Stephan~W Koch, and
  Jerome~V Moloney.
\newblock Modeling and experimental realization of modelocked {VECSEL}
  producing high power sub-\unit[100]{fs} pulses.
\newblock \emph{Appl. Phys. Lett.}, 113\penalty0 (12):\penalty0 121113, 2018.

\bibitem[Meyer et~al.(1995)Meyer, Hoffman, Bartoli, and
  Ram-Mohan]{meyer1995type}
JR~Meyer, CA~Hoffman, FJ~Bartoli, and LR~Ram-Mohan.
\newblock Type-{II} quantum-well lasers for the mid-wavelength infrared.
\newblock \emph{Appl. Phys. Lett.}, 67\penalty0 (6):\penalty0 757--759, 1995.

\bibitem[M{\"o}ller et~al.(2016{\natexlab{a}})M{\"o}ller, Fuchs, Berger,
  Ruiz~Perez, Koch, Hader, Moloney, Koch, and Stolz]{moller2016type}
C~M{\"o}ller, C~Fuchs, C~Berger, A~Ruiz~Perez, M~Koch, Jorg Hader, Jerome~V
  Moloney, Stephan~W Koch, and W~Stolz.
\newblock Type-{II} vertical-external-cavity surface-emitting laser with watt
  level output powers at \unit[1.2]{$\mu$m}.
\newblock \emph{Appl. Phys. Lett.}, 108\penalty0 (7):\penalty0 071102,
  2016{\natexlab{a}}.

\bibitem[M{\"o}ller et~al.(2016{\natexlab{b}})M{\"o}ller, Zhang, Fuchs, Berger,
  Rehn, Perez, Rahimi-Iman, Hader, Koch, Moloney,
  et~al.]{moller2016fundamental}
C~M{\"o}ller, F~Zhang, C~Fuchs, C~Berger, A~Rehn, A~Ruiz Perez, A~Rahimi-Iman,
  Jorg Hader, M~Koch, Jerome~V Moloney, et~al.
\newblock Fundamental transverse mode operation of a type-ii
  vertical-external-cavity surface-emitting laser at 1.2 $\mu$m.
\newblock \emph{Electron. Lett.}, 53\penalty0 (2):\penalty0 93--94,
  2016{\natexlab{b}}.

\bibitem[Rahimi-Iman(2016)]{rahimi2016recent}
Arash Rahimi-Iman.
\newblock Recent advances in vecsels.
\newblock \emph{J. Opt.}, 18\penalty0 (9):\penalty0 093003, 2016.

\bibitem[Tilma et~al.(2015)Tilma, Mangold, Zaugg, Link, Waldburger, Klenner,
  Mayer, Gini, Golling, and Keller]{tilma2015recent}
Bauke~W Tilma, Mario Mangold, Christian~A Zaugg, Sandro~M Link, Dominik
  Waldburger, Alexander Klenner, Aline~S Mayer, Emilio Gini, Matthias Golling,
  and Ursula Keller.
\newblock Recent advances in ultrafast semiconductor disk lasers.
\newblock \emph{Light. Sci. \& Appl.}, 4\penalty0 (7):\penalty0 e310, 2015.

\bibitem[Tropper et~al.(2012)Tropper, Quarterman, and
  Wilcox]{tropper2012ultrafast}
Anne~C. Tropper, Adrian~H. Quarterman, and Keith~G. Wilcox.
\newblock Ultrafast vertical-external-cavity surface-emitting semiconductor
  lasers.
\newblock \emph{Adv. in Semicond. Lasers}, 86:\penalty0 269--300, 2012.

\bibitem[Waldburger et~al.(2016)Waldburger, Link, Mangold, Alfieri, Gini,
  Golling, Tilma, and Keller]{waldburger2016high}
Dominik Waldburger, Sandro~M. Link, Mario Mangold, Cesare G.~E. Alfieri, Emilio
  Gini, Matthias Golling, Bauke~W. Tilma, and Ursula Keller.
\newblock High-power \unit[100]{fs} semiconductor disk lasers.
\newblock \emph{Optica}, 3\penalty0 (8):\penalty0 844--852, 2016.

\bibitem[Wang et~al.(2012)Wang, Heinen, Hader, Dineen, Sparenberg, Weber,
  Kunert, Koch, Moloney, Koch, and Stolz]{LaserPhotonRev12}
T.-L. Wang, B.~Heinen, J.~Hader, C.~Dineen, M.~Sparenberg, A.~Weber, B.~Kunert,
  S.~W. Koch, J.~V. Moloney, M.~Koch, and W.~Stolz.
\newblock Quantum design strategy pushes high-power vertical-external-cavity
  surface-emitting lasers beyond \unit[100]{W}.
\newblock \emph{Laser Photon. Rev.}, 6\penalty0 (5):\penalty0 L12--L14, 2012.

\bibitem[Yang et~al.(2018)Yang, Follman, Albrecht, Heu, Giannini, Cole, and
  Sheik-Bahae]{yang201816}
Z~Yang, D~Follman, AR~Albrecht, P~Heu, N~Giannini, GD~Cole, and M~Sheik-Bahae.
\newblock \unit[16]{W} {DBR}-free membrane semiconductor disk laser with
  dual-{SiC} heatspreader.
\newblock \emph{Electron. Lett.}, 54\penalty0 (7):\penalty0 430--432, 2018.

\bibitem[Zegrya and Andreev(1996)]{zegrya1996theory}
GG~Zegrya and AD~Andreev.
\newblock Theory of the recombination of nonequilibrium carriers in type-ll
  heterostructures.
\newblock \emph{Semicond.}, 2:\penalty0 3, 1996.

\bibitem[Zhang et~al.(2014)Zhang, Heinen, Wichmann, M{\"o}ller, Kunert,
  Rahimi-Iman, Stolz, and Koch]{zhang201423}
Fan Zhang, Bernd Heinen, Matthias Wichmann, Christoph M{\"o}ller, Bernardette
  Kunert, Arash Rahimi-Iman, Wolfgang Stolz, and Martin Koch.
\newblock A 23-watt single-frequency vertical-external-cavity surface-emitting
  laser.
\newblock \emph{Opt. Express}, 22\penalty0 (11):\penalty0 12817--12822, 2014.

\end{thebibliography}


\providecommand{\noopsort}[1]{}\providecommand{\singleletter}[1]{#1}%
\begin{thebibliography}{1}
\providecommand{\natexlab}[1]{#1}
\providecommand{\url}[1]{\texttt{#1}}
\expandafter\ifx\csname urlstyle\endcsname\relax
  \providecommand{\doi}[1]{doi: #1}\else
  \providecommand{\doi}{doi: \begingroup \urlstyle{rm}\Url}\fi

\bibitem[Hader et~al.(2004)Hader, Moloney, and Koch]{hader2004structural}
Jorg Hader, Jerome~V Moloney, and Stephan~W Koch.
\newblock Structural dependence of carrier capture time in semiconductor
  quantum-well lasers.
\newblock \emph{Appl. Phys. Lett.}, 85\penalty0 (3):\penalty0 369--371, 2004.

\end{thebibliography}

\end{document}


\title{Timescales in type-II quantum well carrier recovery}

\author[1]{I.~Kilen}
\author[1,2]{S.~W.~Koch}
\author[1,3]{J.~Hader}
\author[1,3,4]{J.~V.~Moloney}
\affil[1]{College of Optical Sciences, University of Arizona, 1630 East University Boulevard, Tucson, AZ 85721, USA}
\affil[2]{Department of Physics and Material Sciences Center, Philipps-Universit\"at Marburg, Renthof~5, 35032 Marburg, Germany}
\affil[3]{Nonlinear Control Strategies Inc., Tucson, AZ, 85704, USA}
\affil[4]{Department of Mathematics, University of Arizona, 617 N. Santa Rita Ave., Tucson, AZ 85721, USA}
\setcounter{Maxaffil}{0}
\renewcommand\Affilfont{\itshape\small}

\maketitle

Modeling of quantum well (QW) carrier relaxation during mode-locking of vertical external-cavity surface-emitting lasers (VECSELs) requires the solution of the coupled Maxwell semiconductor Bloch equations (MSBE). Here, the electric field is propagated using Maxwell's equation while the SBE describe the microscopic QW carrier dynamics. Together, these form a nonlinear system of differential equations with tens of thousands of variables. To find the mode-locked pulse, we are looking for an asymptotic solution that will appear after the pulse has propagated many thousands of round-trips in the external cavity.

It can be somewhat challenging to extract analytic answers from these models, and sometimes it is interesting to study results from simplified models that can be compared to full simulations to obtain a deeper understanding. Here, we will consider a situation where the mode-locked pulse is already propagating in the cavity. In particular, we will neglect all many-body effects and only focus on the pure carrier relaxation after a pulse has left a QW.

A much-simplified model for the relaxation of the active transition carrier density, $\text{N}(t)$, from the reservoir carrier density, $\text{N}^\text{res}(t)$, can be expressed as,
\begin{equation}
\label{eq:system01}
\begin{aligned}
\frac{\text{d}}{\text{dt}} \mathrm{N}^{\mathrm{res}}(t) &= -\frac{1}{\tau_\text{R}}(\text{N}^\text{res}-F) + \frac{1}{\tau}(\text{N}-\text{N}^\text{res})\\
\frac{\text{d}}{\text{dt}} \mathrm{N}(t) &= -\frac{1}{\tau}(\text{N}-\text{N}^{res}),
\end{aligned}
\end{equation}
where, $\tau_\text{R}$, is the reservoir scattering time, the reservoir relaxes to a constant background density, $\text{F}$, and, $\tau$, is the reservoir to active lasing transition scattering time. This system of equations is linear and has a general solution,
\begin{equation}
\begin{aligned}
\text{N}^\text{res}(t) = F + &A \frac{\gamma-\tau}{2\gamma} \text{e}^{-t\frac{ \tau + 2 \tau_\text{R} + \gamma}{2 \tau \tau_\text{R}}}\left(\text{e}^{\frac{t \gamma}{\tau \tau_\text{R}}}+\frac{\gamma+\tau}{\gamma-\tau}\right)\\
+ &B\frac{\tau_\text{R}}{\gamma} \text{e}^{-t\frac{ \tau + 2 \tau_\text{R} + \gamma}{2 \tau \tau_\text{R}}} \left(\text{e}^{\frac{t \gamma}{\tau \tau_R}} -1\right),
\end{aligned}
\end{equation}
and,
\begin{equation}
\label{eq:modelNSolution}
\begin{aligned}
\text{N}(t)  = F + &A \frac{\tau_\text{R}}{\gamma} \text{e}^{-t\frac{ \tau + 2 \tau_\text{R} + \gamma}{2 \tau \tau_\text{R}}} \left(\text{e}^{\frac{t \gamma}{\tau \tau_R}} -1\right)\\
+& B \frac{\tau + \gamma}{2\gamma} \text{e}^{-t\frac{ \tau + 2 \tau_\text{R} + \gamma}{2 \tau \tau_\text{R}}}\left(\text{e}^{\frac{t \gamma}{\tau \tau_\text{R}}}+\frac{\gamma-\tau}{\gamma + \tau}\right),
\end{aligned}
\end{equation}
where $\gamma = \sqrt{\tau^2 + 4 \tau_\text{R}^2}$. A and B are arbitrary constants that can be used with initial conditions or boundary conditions to determine the full solution. However, we will not need these because we can already see that the dynamics of, N(t), is dictated by sums and differences of two different exponentials, $\text{exp}\left(-t\frac{\tau + 2\tau_\text{R} \pm \gamma}{2\tau \tau_\text{R}}\right)$. Denote the two timescales as, 
\begin{equation}
\label{eq:doubleExpTimescale}
\text{T}_\pm = \frac{2 \tau \tau_\text{R}}{\tau + 2 \tau_\text{R} \pm \gamma}.
\end{equation}
It is evident that the carriers in the lasing transition have a two stage relaxation process where one time is fast ($\text{T}_+$), and one is slow ($\text{T}_-$). It is worth pointing out that neither one of these timescales are equal to the scattering timescales, $\tau$, or, $\tau_\text{R}$. Indeed, we will see that these times can be very different.

In general, we can simplify, $\text{T}_\pm$, by comparing it to the reservoir recovery time and notice that if we want, $\text{T}_\pm = \alpha \tau_R$, then we must also have,
\begin{equation}
\label{eq:tau_alpha}
\tau = \frac{\alpha(\alpha-2)}{\alpha-1}\tau_\text{R},
\end{equation}
where $\tau>0$ only for $\alpha \in (0,1)\cup(2,\infty)$. Fig.~\ref{fig:solutions} shows an overview of the solution space. 

\begin{figure}[ht]
\centerline{\includegraphics[width=0.5\linewidth]{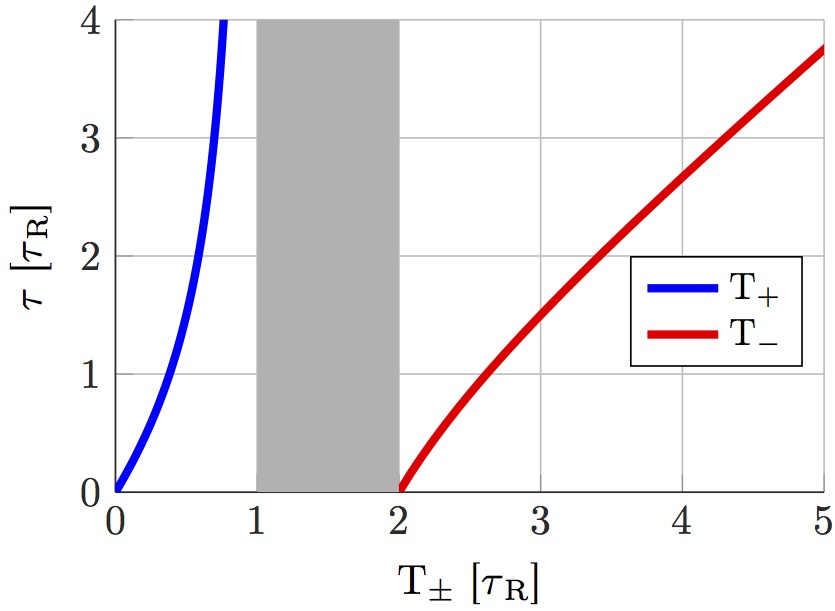}}
\caption{A map of carrier recovery times, $\text{T}_\pm$, of the active lasing transition as parameterized by the scattering time, $\tau$, from Eq.~\ref{eq:tau_alpha}. All times are in units of the reservoir scattering time, $\tau_\text{R}$. The grey region contains no solutions.}
\label{fig:solutions} 
\end{figure}
A surprising observation in Fig.~\ref{fig:solutions} is the fast recovery time, $\text{T}_+ <\tau_\text{R}$. The significance of this upper bound is that no matter how large the scattering time, $\tau$, becomes, there are carriers recovering into the lasing transition state on the timescale, $\text{T}_+ <\tau_\text{R}$.  Also, the lower boundary is zero, which is not possible because both, $\tau$, and, $\tau_\text{R}$, are nonzero. This fast relaxation can become troublesome for stable mode-locking because it can cause too many carriers to recover directly after the main pulse has passed. In this case, a tail can begin to form behind the primary pulse.

Fig.~\ref{fig:solutions} shows that the timescale, $\text{T}_+$, varies between zero and up to the reservoir scattering timescale, $\tau_\text{R}$, which is usually much longer than the duration of a mode-locked pulse. Thus, one can always design the system with a sufficiently long carrier recovery time that prevents pulse instabilities due to too many unsaturated carriers. 

We will compare the QW carrier densities computed from mode-locked simulations using the full MSBE with fits to double exponential times computed with the simple analytic model. The fit function will be similar to Eq.~\ref{eq:modelNSolution} with $f(t) = a + b \text{e}^{-t/\text{T}_+} + c \text{e}^{-t/\text{T}_-}$ where, $\text{T}_\pm$, are the timescales computed in Eq.~\ref{eq:doubleExpTimescale} and (a,b,c) are fit parameters. As an example, we take the electron and hole QW carrier densities in a simulation where $\tau_\text{R} = 30\,\text{ps}$, $\tau_e = 1\,\text{ps}$, and $\tau_h = 5\,\text{ps}$. For these parameters we get, $\text{T}^\text{e}_+ = 0.496\,\text{ps}$, and, $\text{T}^\text{e}_- = 60.5\,\text{ps}$, for the electrons by using Eq.~\ref{eq:doubleExpTimescale}. For the holes we similarly get, $\text{T}^\text{h}_+ = 2.396\,\text{ps}$, and, $\text{T}^\text{h}_- = 62.6\,\text{ps}$.

In Fig.~\ref{fig:relaxFit}ab) we see the electron and hole carrier recovery from a single QW as simulated using the MSBE with {type-II} QWs. A mode-locked pulse is propagating in the cavity and periodically extracts carriers form the QW. To fit this data with the simpler model, we focus on the recovery time between two successive pulse interactions. Fig.~\ref{fig:relaxFit} shows that there is an excellent agreement between the timescales computed with the analytic model and the full MSBE. The fit parameters were found with 95\% confidence bounds to be within $(a,b,c) = (1.802992\pm 0.000014,\,-57116\pm 49,\,-0.079754\pm 0.000037)$ for the electron density and $(a,b,c) = (1.801160\pm 0.000027,\,-0.41736\pm 0.00030,\,-0.087001\pm 0.000037)$ for the hole density.

In summary, we derived a simple model of the {type-II} QW carrier relaxation. Showed how the parameters that go into the SBE, $\tau_\text{scatt}$, $\tau_e$, and, $\tau_h$, are related to the replenishing timescales for the active lasing transition through an exponential relaxations with two characteristic timescales, $\text{T}_\pm$. The carrier dynamics caused by these timescales should not be confused with other microscopic many-body dynamics from, e.g., carrier scattering, which has a characteristic timescale around \unit[100]{fs}. Indeed, the two timescales, $\text{T}_\pm$, were derived under the assumption of a pure carrier relaxation with no carrier scattering, no electric field, or any nonlinear influences. Finally, we verify that the full simulation of the MSBE with the two stage pump relaxation produces a carrier recovery that agrees with the simple analytic result in the absence of an electric field.

This material is based upon work supported by the Air Force Office of Scientific Research under award numbers FA9550-17-1-0246. The Marburg research is supported by the Deutsche Forschungsgemeinschaft via the Sonderforschungsbereich 1083. We thank Wolfgang Stolz for insightful discussions.
\begin{figure}[ht]
\centerline{\includegraphics[width=0.5\linewidth]{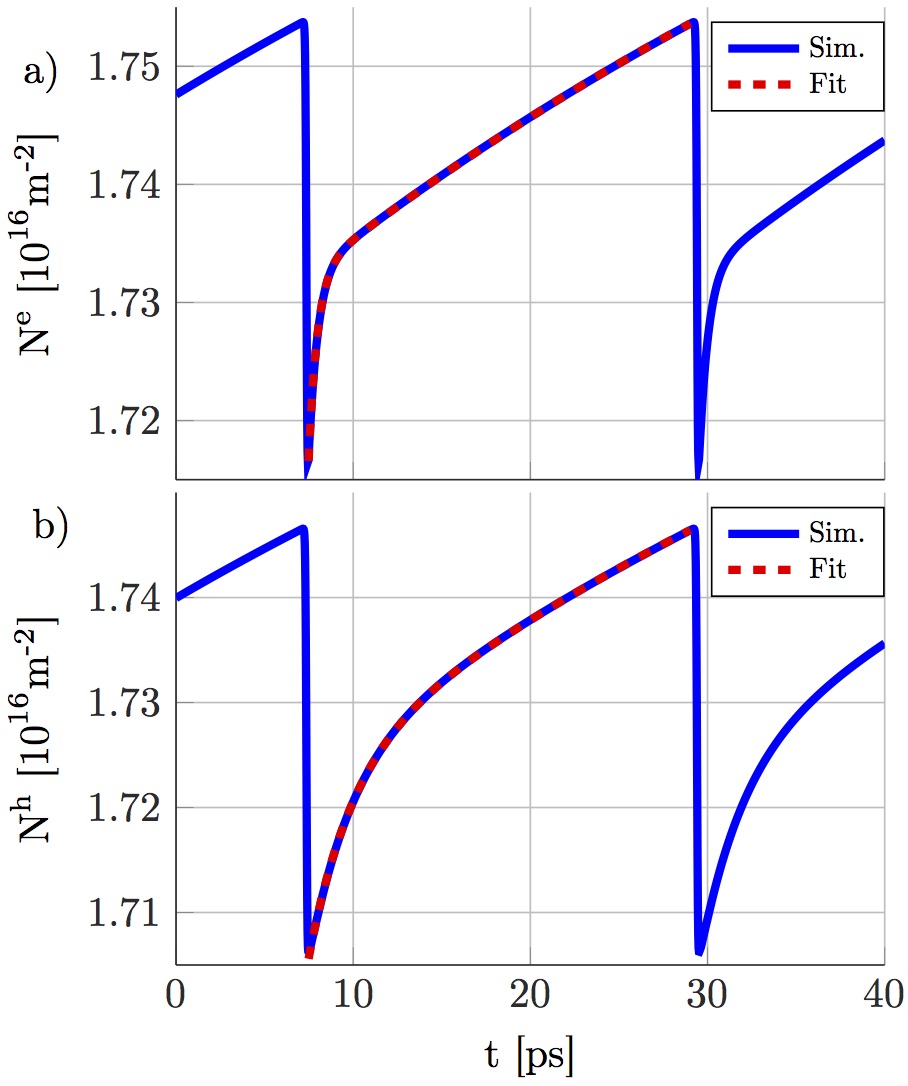}}
\caption{A snapshot of the QW carrier density in the active lasing electrons (a) and holes (b) from a mode-locked simulation using the full MSBE (solid blue) and a double exponential fit (dashed red). We are using the parameters $\tau_\text{R} = 30\,\text{ps}$, $\tau_e = 1\,\text{ps}$, and $\tau_h = 5\,\text{ps}$.}
\label{fig:relaxFit} 
\end{figure}


\title{Non-equilibrium pulse instability}

\author[1]{I.~Kilen}
\author[1,2]{S.~W.~Koch}
\author[1,3]{J.~Hader}
\author[1,3,4]{J.~V.~Moloney}
\affil[1]{College of Optical Sciences, University of Arizona, 1630 East University Boulevard, Tucson, AZ 85721, USA}
\affil[2]{Department of Physics and Material Sciences Center, Philipps-Universit\"at Marburg, Renthof~5, 35032 Marburg, Germany}
\affil[3]{Nonlinear Control Strategies Inc., Tucson, AZ, 85704, USA}
\affil[4]{Department of Mathematics, University of Arizona, 617 N. Santa Rita Ave., Tucson, AZ 85721, USA}
\setcounter{Maxaffil}{0}
\renewcommand\Affilfont{\itshape\small}

\maketitle

In the process of exploring mode-locked pulses with type-II quantum wells (QWs), we find a true non-equilibrium pulse instability for fast electron or hole relaxation times. This instability does not produce multiple pulses but is a cyclic replenishing of the mode-locked pulse tail that overtakes the primary pulse. Here, we present some details on this instability and find the cause to be a combination of fast recovery times and too high background carrier density. Too much pumping is known to cause multiple pulsing in type-I QWs because it creates too many unsaturated carriers that can be taken advantage of by a second pulse. In this appendix, we will observe the influence of this fast relaxation time on the mode-locked pulse. We do not expect this to be a significant problem in mode-locking type-II QWs, because we will see that it is easily avoidable by tuning the background density and the relevant rates are very short and might be impossible to obtain in realizations of gain chips with type-II QWs. However, this appendix is useful as an example of the dynamics of this mode-locked pulse instability.

\begin{figure}[ht]
\centerline{\includegraphics[width=0.6\linewidth]{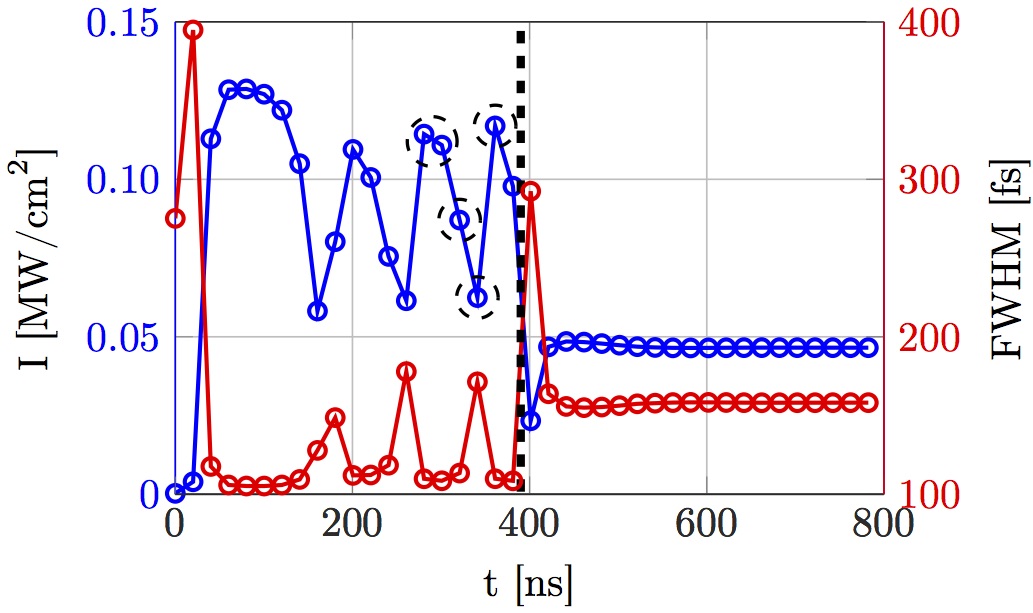}}
\caption{The transient peak intensity and FWHM of a pulse propagating in a VECSEL cavity with $\tau_e=1\,\text{ps}$, $\tau_h=1\,\text{ps}$, and $\tau_\text{scatt}=30\,\text{ps}$. The vertical dashed line indicates the time when the background carrier density is reduced by \unit[5]{\%}. The dashed circles indicates the location of pulses used in Fig.~\ref{fig:pulseInstab}.}
\label{fig:reduceDensity} 
\end{figure}
In appendix A, we used a simple model to derive the relaxation timescales for the carrier density in the lasing transition. We found that the carrier densities relax on a fast and a slow timescale, $\text{T}_\pm$,  that are nonlinear combinations of the reservoir recovery time, $\tau_\text{scatt}$, and the electron and hole recovery times, $\tau_\text{e}$, and, $\tau_\text{h}$. In particular, if $\tau_\text{e} \ll \tau_\text{scatt}$ then the fast timescale, $\text{T}^\text{e}_+$, can become faster than, $\tau_\text{e}$.  

In Fig.~\ref{fig:reduceDensity}, we can see the peak intensity and full-width half-maximum (FWHM) of a pulse propagating in a VECSEL cavity with a round trip time of about \unit[21]{ps}. The initial pulse is a low energy seed pulse with an FWHM of \unit[270]{fs}. In this simulation, the initial background carrier density of the reservoir is \unit[$1.82\cdot10^{16}$]{$\mathrm{m}^{-2}$}, and the reservoir is recovering with a timescale $\tau_\text{scatt} = 30\,\text{ps}$. The electrons and holes in the active lasing transition are both recovering with a \unit[1]{ps} timescale. During the first \unit[60]{ns}, the seed pulse is increasing in energy and decreasing in FWHM until it becomes strong enough to saturate the gain chip QWs. However, there is no convergence to a stable mode-locked pulse. Instead, the peak amplitude and FHWM fluctuate with no sign of slowing down. Fig.~\ref{fig:pulseInstab} gives a more detailed look at the pulse instability.

The pulse instability seen in Fig.~\ref{fig:reduceDensity} comes from the short electron and hole relaxation times used for this example. We selected this example for its clarity. However, similarly unstable pulses appear for other relaxation times close to \unit[1]{ps} relaxation time. In particular, the pulses become unstable if only one of the two timescales is short, e.g. the electron relaxation time is \unit[1]{ps}, while the hole relaxation time can be \unit[40]{ps} or longer. When the carrier relaxation time is too short, it can cause too many carriers to recover after or even during the pulse interaction with the propagating pulse. A natural way to verify this is to reduce the background pump density, which should stabilize the pulse. The pump density is reduced by \unit[5]{\%} at \unit[390]{ns} which is denoted by a dashed black line in Fig.~\ref{fig:reduceDensity}. After this, the pulse peak intensity and FWHM go through a short equilibration phase and quickly stabilizes.

Fig.~\ref{fig:pulseInstab} contain snapshots of the unstable pulse envelope and spectrum evolving over about 3600 round-trips. Figs.~\ref{fig:pulseInstab}a-e) depict the pulse envelope and Figs.~\ref{fig:pulseInstab}f-j) shows the computed pulse spectrum. Here, it is clear that a second pulse is growing from the tail of the original pulse. The new pulse spectra appear at a higher energy than the original pulse spectrum. As the second pulse emerges, the initial pulse spectrum is pushed into a lower energy region and is now experiencing loss of energy as the emerging pulse continues to grow. After a few more round-trips, there is only one pulse remaining, and another pulse tail has developed. The final pulse state is very similar to the initial pulse in these panels. This cycle will continue if this simulation is not interrupted.

\begin{figure}[ht]
\centerline{\includegraphics[width=0.6\linewidth]{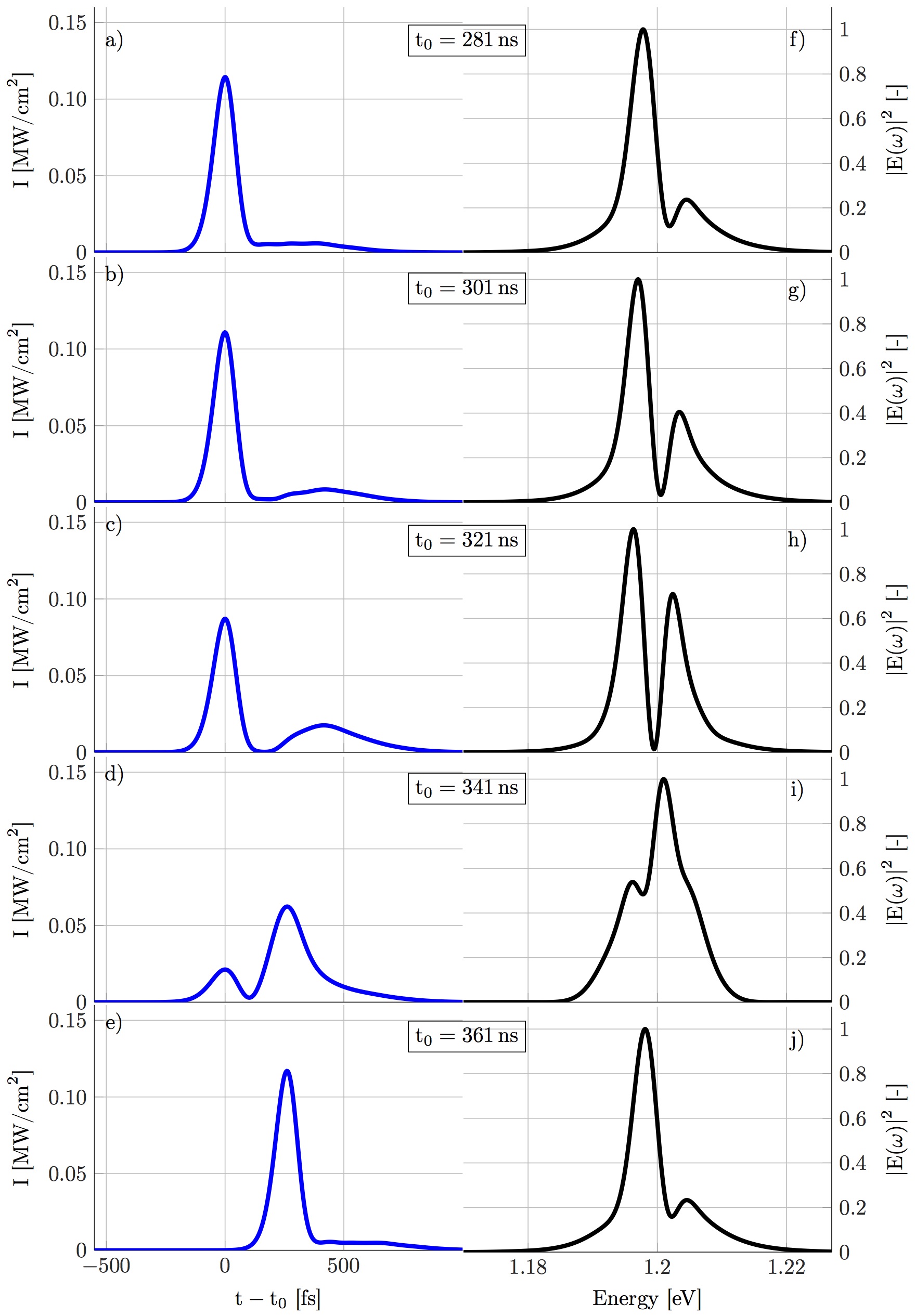}}
\caption{The evolution of an unstable pulse. Column 1 (Panels: a-e) show the pulse envelope at the given time intervals, while column 2 (Panels: f-j) is the computed spectrum from the pulse in column 1. The sample times are also marked in Fig.~\ref{fig:reduceDensity} as dashed circles. The time axis in panels a-d) is centered relative to the first pulse peak denoted by, $\mathrm{t}_0$. In panel e), the time axis is translated to align the pulse peak relative to the second pulse in panel d).}
\label{fig:pulseInstab} 
\end{figure}

In summary, a too fast Type II QW electron/hole recovery time with too high background pump density produces an unstable mode-locked pulse. However, reducing the background density will stabilize the mode-locked pulse, which means that experiments can easily avoid this situation. The instability is a product of the current VECSEL setup, and it might be possible to design a VECSEL where these faster relaxation times are an advantage, e.g., if the goal is to produce high energy pulses.
 
This material is based upon work supported by the Air Force Office of Scientific Research under award numbers FA9550-17-1-0246. The Marburg research is supported by the Deutsche Forschungsgemeinschaft via the Sonderforschungsbereich 1083. We thank Wolfgang Stolz for insightful discussions.


\title{Type-II quantum well carrier capture dynamics}

\author[1]{I.~Kilen}
\author[1,2]{S.~W.~Koch}
\author[1,3]{J.~Hader}
\author[1,3,4]{J.~V.~Moloney}
\affil[1]{College of Optical Sciences, University of Arizona, 1630 East University Boulevard, Tucson, AZ 85721, USA}
\affil[2]{Department of Physics and Material Sciences Center, Philipps-Universit\"at Marburg, Renthof~5, 35032 Marburg, Germany}
\affil[3]{Nonlinear Control Strategies Inc., Tucson, AZ, 85704, USA}
\affil[4]{Department of Mathematics, University of Arizona, 617 N. Santa Rita Ave., Tucson, AZ 85721, USA}
\setcounter{Maxaffil}{0}
\renewcommand\Affilfont{\itshape\small}

\maketitle

\begin{figure}[ht]
\centerline{\includegraphics[width=0.6\linewidth]{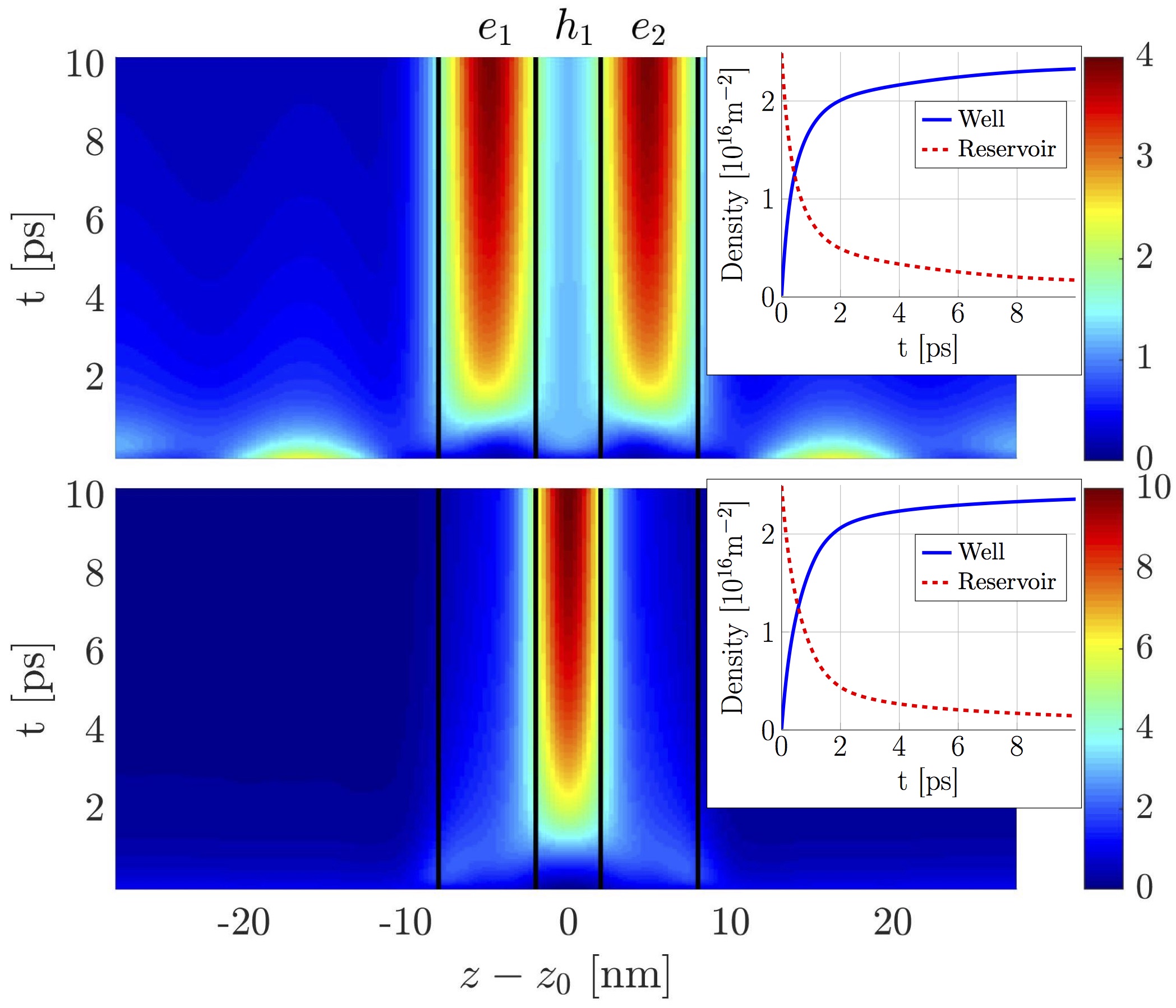}}
\caption{Fully microscopically calculated carrier relaxation of electrons/holes (top/bottom) in a type-II W QW. Shown in color is the charge distribution in arbitrary units as a function of time. Initially, at $t=0\,\text{ps}$, a density of \unit[$2.5\cdot 10^{16}$]{$\text{m}^{-2}$} is placed exclusively in states of the reservoir, energetically above the quantum well, while the states of the well are empty. $e_1$ and $e_2$ mark the positions of the electron wells and $h_1$ marks the position of the hole well. The inset shows the total density in reservoir states (dashed red) and states located in the well (solid blue).}
\label{fig:modeLockedExample}
\end{figure}

To get realistic estimates for the capture times, we ran fully microscopic multiband many-body scattering simulations as shown for type-I quantum well systems in Ref.\cite{hader2004structural}. For type-II W-wells these simulations are numerically extremely demanding due to the rather high number of subbands that are confined in the well. Also, in the mode-locked case the changes in the distributions and resulting densities are rather small which requires a prohibitively high numerical precision in order to track over the relevant timescales as well as the source- and drain-dynamics. As an example of the carrier capture process, we show the case where a number of carriers as typical for lasing operation is initially placed exclusively into states localized in the barrier. Then we monitor the change in density within the lowest subbands confined in the well that contribute to lasing. As shown in Fig.~\ref{fig:modeLockedExample}, the initial change in density in the lowest bands is very fast since the final states of the scattering processes are empty and the initial states have high occupation probabilities. A situation comparable to the mode-locked state is found after a few picoseconds where the densities and occupations of the lasing states are near the inversion level. Here we find timescales on the order of ten picoseconds with the exact times depending on details of the carrier distributions.

This material is based upon work supported by the Air Force Office of Scientific Research under award numbers FA9550-17-1-0246. The Marburg research is supported by the Deutsche Forschungsgemeinschaft via the Sonderforschungsbereich 1083. We thank Wolfgang Stolz for insightful discussions.

\bibliographystyle{plainnat}
\bibliography{supmatC}